\documentclass[aps,prb,twocolumn,superscriptaddress, showpacs,preprintnumbers,amsmath,amssymb]{revtex4-1}

\usepackage{xspace}
\usepackage[active]{srcltx}
\usepackage{amsmath,amssymb}
\usepackage{graphicx}
\usepackage[]{natbib}
\usepackage{longtable}
\usepackage{hyperref}
\hypersetup{
    unicode=false,          
    pdftoolbar=true,        
    pdfmenubar=true,        
    pdffitwindow=false,     
    pdfstartview={FitH},    
    pdftitle={My title},    
    pdfauthor={Author},     
    pdfsubject={Subject},   
    pdfcreator={Creator},   
    pdfproducer={Producer}, 
    pdfkeywords={keyword1} {key2} {key3}, 
    pdfnewwindow=true,      
    colorlinks=true,       
    linkcolor=blue,          
    citecolor=blue,        
    filecolor=blue,      
    urlcolor=blue,       
    plainpages=false        
}
\def\cm{cm$^{-1}$}
\def\etbr{$\kappa$-(BEDT\--TTF)$_2$\-Cu\-[N(CN)$_{2}$]Br}
\def\etcl{$\kappa$-(BEDT\--TTF)$_2$\-Cu\-[N(CN)$_{2}$]Cl}

\begin{document}

\title{Antiferromagnetic fluctuations in the quasi-two-dimensional organic superconductor  detected by Raman spectroscopy.}

\author{Natalia~Drichko}\email{Corresponding author: drichko@pha.jhu.edu} \affiliation{Dept. of Physics and Astronomy, Johns Hopkins University,  Baltimore MD, USA}
\author{Rudi Hackl}
\affiliation{Walther Meissner Institut, Bayerische Akademie der Wissenschaften, 85748 Garching, Germany}
\author{John A. Schlueter}
\affiliation{Materials Science Division, Argonne National Laboratory, Argonne, IL 60439, USA; NSF,  Arlington, Virginia 22230, USA  }

\date{\today}

\begin{abstract}
  Using Raman scattering, the quasi-two dimensional organic superconductor \etbr\ ($T_c=11.8$\,K) and the related antiferromagnet \etcl\ are studied. Raman scattering provides unique spectroscopic information about magnetic degrees of freedom that has been otherwise unavailable on such organic conductors. Below $T=200$\,K  a broad band at about 500~\cm\  develops in both compounds. We identify this band with  two-magnon excitation. The position and the temperature dependence of the spectral weight are similar in the antiferromagnet   and in the metallic Fermi-liquid. We conclude that antiferromagnetic correlations are similarly present in the magnetic insulator and the Fermi-liquid state of the superconductor.
\end{abstract}


\maketitle

\section{Introduction}

In correlated electron systems, metallic and magnetically ordered phases are typically in close proximity and can be controlled by a nonthermal parameter  such as doping, pressure or magnetic field. A doping-controlled transition produces such important metallic and superconducting compounds as copper-oxide\cite{Lee:2006} and iron-based\cite{Scalapino:2012} superconductors. A pressure-controlled transition from a theoretical point of view results in a variation of $U/t$, the ratio of the Coulomb over kinetic energy where the latter is represented by the hopping integral. The ratio $U/t$ can be changed by modifying bandwidth $W\sim t$. Studies of the bandwidth-controlled transition  are aimed on elucidating quantum critical behaviour  and non-Fermi liquid state found close to the antiferromagnetic  (AF) state\cite{Gegenwart2008,Coleman2010}. Most of the  systems where this transition is studied possess orbital degrees of freedom, which adds complexity to the physics, and the transition could be selective for particular orbitals\cite{Simonson2012}. A rare example of a bandwidth-controlled phase transition from AF insulator to a Fermi-liquid metal where orbital physics is irrelevant is presented by the family of quasi-two-dimensional (q2D) BEDT-TTF\footnote{BEDT-TTF=ET= bis(ethylene-dithio)-tetrathiafulvalene}-based organic conductors (see Fig.~1).



\begin{figure}
\begin{center}
  \includegraphics[width=9cm]{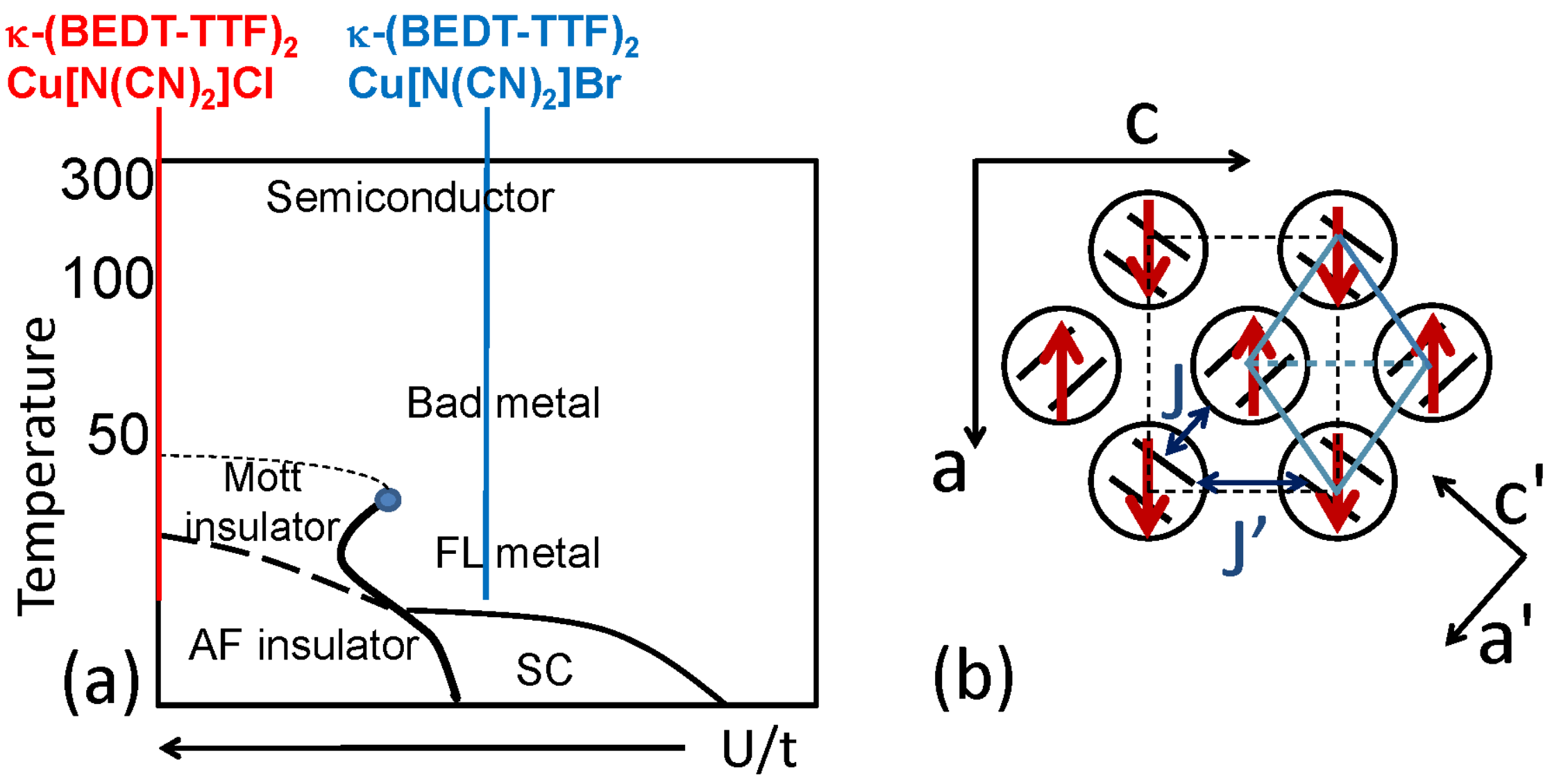}\\
  \caption{(a) Schematic phase diagram for BEDT-TTF-based materials at half filling. The studied compounds are indicated by a red an a blue vertical line. (b) Schematic structure of the $\kappa$-(BEDT-TTF)$_2$Cu[N(CN)$_2$]$X$ ($X$=Cl,~Br) layer. The (BEDT-TTF)$_2^{+1}$ dimers are encircled. The putative orientation of the spins is indicated in red. The magnetic unit cell is given in blue. Due to the orthorhombic structure of the crystallographic unit cell (black dashes), having dimensions $a$ and $c$ along the unit vectors $\hat a$ and $\hat c$, the magnetic unit cell is characterized by nearly equilateral triangles. The axes $\hat a^\prime = 1/\sqrt{2}(\hat a + \hat c)$ and $\hat c^\prime = 1/\sqrt{2}(\hat c - \hat a)$ are considered approximately parallel (for theoretical purposes) to the axes of the magnetic unit cell. The hopping integral $t$ ($t^\prime$) and the exchange coupling $J$ ($J^\prime$) are oriented along the full (broken) blue line.
  }
  \label{fig:PD}
\end{center}
\end{figure}

In organic conductors  bandwidth $W\sim t$ is varied by moderate pressures or by small modifications of the anions. Fig.~\ref{fig:PD}\,(a) shows that the entire phase diagram of prototypical BEDT-TTF-based compounds can be accessed with \etcl\ ($\kappa$-ET-Cl), \etbr\ ($\kappa$-ET-Br), and application of pressure.
Antiferromagnetic (AF) order  is observed below 23\,K in the Mott insulator $\kappa$-ET-Cl\cite{Miyagawa:1995,Antal2009}.  The increase of the bandwidth by substituting Br for Cl goes along with a decrease of $U/t$ and yields the superconductor $\kappa$-ET-Br with $T_c=11.8$\,K. Below $T_c$ AF spin fluctuations were suggested to be at the origin of potentially unconventional superconductivity \cite{Ishiguro1998,Toyota:2007,Powell06}. Above $T_c$ and below 40\,K the resistivity shows typical Fermi-liquid (FL) behavior\cite{Strack2005}. However, the decrease of the spin-lattice relaxation rate on cooling as observed by NMR \cite{Kawamoto1995,Wzietek1996} is inconsistent with the standard FL picture. Possible explanations include the development of a  pseudogap or a mixed metal-insulator state\cite{Yusuf2007,Powell09}.

The aim of our work was to investigate magnetic excitations in the FL state of the organic superconductor $\kappa$-ET-Br (T$_c$=11.8~K) in order to clarify the nature of AF fluctuations in this state. This is important both for understanding of the nature of this AF to Fermi-liquid transition, and for  understanding of unconventional superconductivity in this material\cite{Ishiguro1998,Toyota:2007,Powell06}. Investigation of the magnetic excitations in the $\kappa$-ET-Cl and $\kappa$-ET-Br is also a step to understanding the  gapless spin liquid candidates among q2D organic conductors\cite{Yamashita2008,Pratt2011,Yamashita2010,Yamashita2011,Isono2014}.
To study magnetic excitations in $\kappa$-ET-Br and $\kappa$-ET-Cl we used  Raman scattering technique. It is the only spectroscopic technique which can access magnetic excitations in organic conductors, since neutron scattering studies on these materials are difficult. Raman scattering is known to provide access to spin excitations \cite{Cottam,Devereaux:2007}. In a recent light scattering experiment the existence of spin excitations was indeed reported for $\kappa$-ET-Cl and signatures of two-magnon excitations were observed in the range of 500\,cm$^{-1}$\cite{Nakamura2014}.

In this letter, we study $\kappa$-ET-Br and address the question whether the spin fluctuations persist into the metallic state and have enough energy to explain the anomalies in the normal state and potentially serve as an exchange boson for unconventional superconductivity. Our results show, that AF correlations similar  to those in $\kappa$-ET-Cl exist in  the FL  state of  $\kappa$-ET-Br. The polarization dependence of the signal is consistent with checkerboard AF order in (ac) plane [see Fig.~\ref{fig:PD}\,(b)] similar to that in the cuprates.


Crystals of $\kappa$-ET-Br and $\kappa$-ET-Cl have layered structures, where layers of cations and anions are parallel to the $(ac)$ crystallographic planes and alternate along $b$-axis. They are iso-structural and have an  orthorhombic $Pnma$ symmetry. The insulating and non-magnetic anion layers Cu[N(CN)$_2$]$X$ ($X=$\,Cl, Br) serve as charge reservoirs whereas the conductive cation layers define the physical properties. Only $\pi$ molecular orbitals  are involved in the formation of the conduction band and in magnetic interactions. This leads to a relatively simple electronic structure. The cation layer is well approximated  by  a two-dimensional (2D) triangular lattice of  dimers of (BEDT-TTF)$_2^{+1}$ with spin 1/2 and direct spin exchange $J$ as shown in  Fig.~\ref{fig:PD}\,(b).

The Raman spectra of $\kappa$-ET-Br and $\kappa$-ET-Cl were measured using a calibrated Raman setup equipped with a Jarrell-Ash 25-100 double monochromator. For excitation, the Ar$^+$ laser line at 514.5\,nm was used. The samples were cooled down in a He-flow cryostat. The cooling rates were in the range from 0.2\,K/min to 1\,K/min so as to avoid a mixed state of metallic and insulating domains in  $\kappa$-ET-Br  \cite{Yoneyama2005,Fournier2007}. The absence of luminescence at 3500~\cm\ proved that $\kappa$-ET-Br is in a homogeneous metallic state\cite{Drichko:2013a}.

The polarization vectors of the incident and scattered photons ${\bf e}_{i,s}$ were parallel to the $ac$ plane, with the light propagating  along the $b$-axis, ${\bf k}_{i,s}||\pm b$. The experimental data below are obtained  in  $b(\hat c \hat a)\bar{b}$ and $b(\hat c^\prime \hat a^\prime)\bar{b}$ polarization, where $\hat a^\prime = 1/\sqrt{2}(\hat a + \hat c)$ and $\hat c^\prime = 1/\sqrt{2}(\hat c - \hat a)$. $\hat c$ and  $\hat a$  is the coordinate system of the crystal, whereas that of the magnetic unit cell is approximately oriented along $\hat a^\prime,~\hat c^\prime$ (see Fig.~1). We can thus project $b(\hat c \hat a)\bar{b}$ and $b(\hat c^\prime \hat a^\prime)\bar{b}$ respectively on $B_{2g}$ and $B_{1g}$ symmetry of a square lattice, where $B_{1g}$ symmetry corresponds approximately to the magnetic $x^\prime y^\prime$ polarization used for square 2D antiferromagnets \cite{Fleury:1968,Devereaux:2007}. Since $a$ and $c$ are substantially different the approximation is poor. It is nevertheless useful for the analysis of the magnetic scattering since $J^\prime\approx 0.2J$ for $\kappa$-ET-Cl \cite{Mori2002,Nakamura2014} and one is far away from frustration ($J^\prime\approx J$).

\begin{figure}
  \includegraphics[width=10cm]{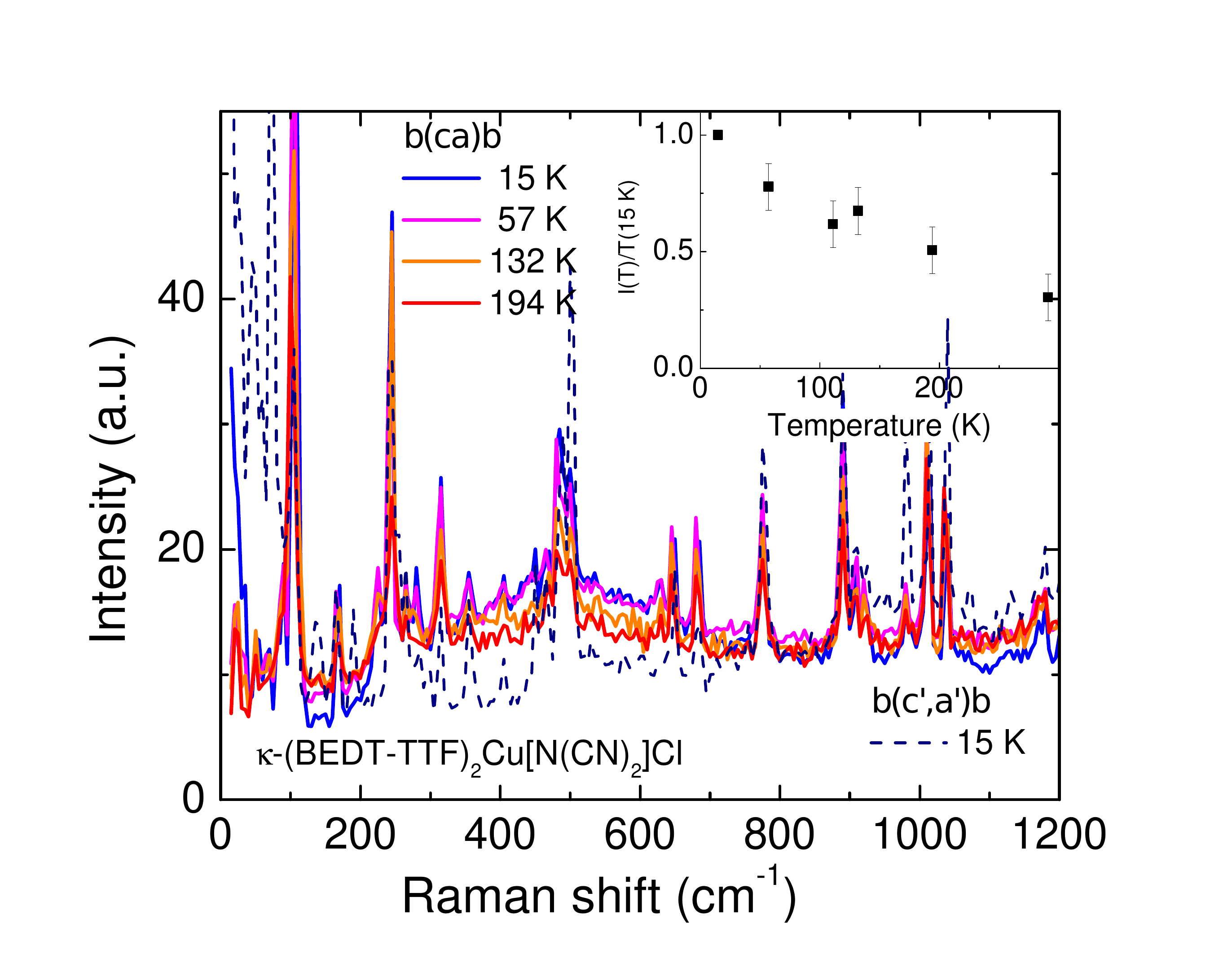}\\
  \caption{Temperature dependence of Raman spectra of \etcl\ in $b(\hat c\hat a)\bar{b}$  polarization  at temperatures between 15 and 200 K (solid lines). A dashed line shows a spectrum in  $b(\hat c^\prime \hat a^\prime)\bar{b}$ polarization at 15\,K. The inset shows a temperature dependence of the spectral weight of the two-magnon excitations, estimated by $I(T)/I(15\,K)$.}
  \label{Cl}
\end{figure}

\begin{figure}
\begin{center}
  \includegraphics[width=10cm]{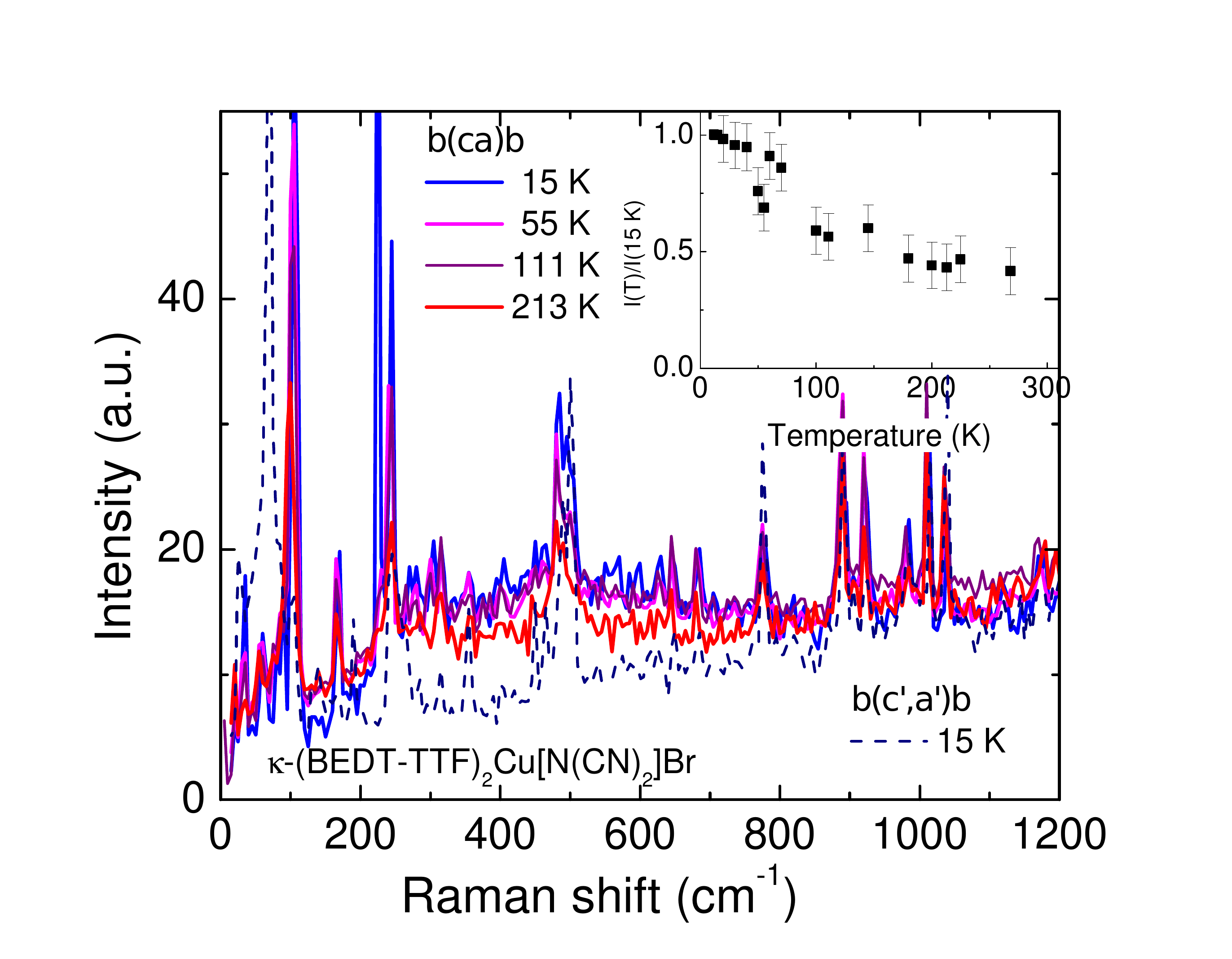}\\
  \caption{Temperature dependence of Raman spectra of \etbr\ in $b(\hat c\hat a)\bar{b}$  polarization  at temperatures between 15 and 213 K, spectra are shown with solid lines. A dashed line shows a spectrum in  $b(\hat c^\prime \hat a^\prime)\bar{b}$  polarization at 15\,K, note the absence of two-magnon maximum at 500~\cm. The inset shows a temperature dependence of the spectral weight of the two-magnon excitations, estimated by $I(T)/I(15 K)$}\label{BrND}
\end{center}
\end{figure}

The Raman spectra of $\kappa$-ET-Cl and $\kappa$-ET-Br are shown in Figs.~\ref{Cl} and \ref{BrND}, respectively. They are characterized by a broad continuum and the expected large number of narrow vibrational excitations. All the lines at frequencies above approximately 50\,\cm\ have been assigned to BEDT-TTF molecular vibrations\cite{Kozlov1987}. At lower frequencies one finds vibrations of the lattice \cite{Brillante:1997}. We focus now on the continuum part of which originates in magnetic excitations.

In $\kappa$-ET-Cl in $b(\hat c\hat a)b$ we observe a wide band with the center at approximately 500\,\cm\ developing below 200\,K (see Fig. \ref{Cl}).  The observed maximum is quite symmetric, and a Gaussian fit at 15\,K yields a maximum frequency of 480\,\cm\ and a width (FWHM) of 400\,\cm.  This response is  absent in $b(\hat a\hat a)\bar{b}$, $b(\hat c\hat c)\bar{b}$ polarizations, and in $b(\hat c^\prime \hat a^\prime)\bar{b}$ (see the spectrum shown with the dashed line in Fig.~\ref{Cl}).

We follow the development of this wide peak on cooling by  calculating its relative spectral weight at  temperature $T$ as $I(T)/I(15\,K)$, see inset in Fig.~\ref{Cl}. Here
\begin{equation}
  \label{eq:int}
  I( T)=\int_{\rm 200\,cm^{-1}}^{\rm 800\,cm^{-1}}(\chi_\mu^{\prime\prime}(\omega, T)- C_{\rm background})d\omega,
\end{equation}
with $\chi_\mu^{\prime\prime}(\omega, T)$ is Raman response at symmetry $\mu=B_{1g} ~B_{2g}$. The response is obtained by dividing the experimental cross section by the Bose-Einstein factor $\{1+n(T,\Omega)\}=[1-\exp(-\hbar\Omega/k_BT)]^{-1}$. $C_{\rm background}$ is a continuum resulting from electronic excitations and wide-range luminescence, which does not show substantial polarization and temperature dependence in the studied frequency range, and in which we are not interested here. For subtracting it we use the $b(\hat c^\prime \hat a^\prime)\bar{b}$ spectra that depend fairly linearly on frequency and approximate them by an analytic function  $C_{\rm background} = C_0(200~cm^{-1})+ C_1*\omega$, where $C_0$ and $C_1$ are constants \footnote{This allowed us to use the data for $\kappa$-ET-Br measured at two different Raman setups, see Supplemental Materials}. After subtracting the linear background from the $b(\hat c\hat a)\bar{b}$ spectra we can isolate the maximum in $b(\hat c^\prime \hat a^\prime)\bar{b}$ polarization.
Its spectral weight increases continuously from room temperature down to 15~K (inset of Fig.~\ref{Cl}).

In the Raman spectra of $\kappa$-ET-Br we observe  a similar wide band in $b(\hat c\hat a)\bar{b}$ polarization. The spectral weight of this band also increases on cooling (see the inset in Fig. \ref{BrND}) all the way down to 15~K.

The wide band at 500~\cm\ which appears in the $b(\hat c\hat a)\bar{b}$ polarization spectra in $\kappa$-ET-Br and $\kappa$-ET-Cl does not coincide in energies with any of the electronic excitations observed in these materials, for instance interband excitations and a transition between Hubbard bands observed at  2000-4000~\cm, and a Drude peak at frequencies below 200\,\cm(Ref.\onlinecite{Faltermeier07}). On the other hand, we can expect signatures of two-magnon scattering in this frequency range and polarization \cite{Nakamura2014}.

According to  angle-dependent NMR measurements in the ordered state of $\kappa$-ET-Cl the magnetic moments show AF order in $(ac)$ planes, with an  easy axis parallel to the $b$ axis of the crystal (perpendicular to  BEDT-TTF layers)\cite{Smith2003,Antal2009}. To the best of our knowledge there is no information about the symmetry of AF order. Two possibilities have been discussed\cite{Mori2002},  one is a checkerboard order (Fig.~\ref{fig:PD} (b)) where spins from the two different magnetic sublattices  occupy two different symmetry sites, another is an order where AF-arranged chains are parallel to the $a$ and $c$ axes. In an approximation that takes into account only nearest-neighbor spin exchange, justified by the relatively small value of the next-nearest-neighbor interaction $J^\prime=0.2J$, the spin pattern depends on a ratio between $J/J^\prime$ (see Fig.~~\ref{fig:PD}), and checkerboard order was considered to be more probable for $\kappa$-ET-Cu-Cl\cite{Mori2002}.

The general form of an exchange magnetic Raman scattering operator for  AF is $\hat{R}=\boldsymbol{S}_i J_{ij} \boldsymbol{S}_j (\boldsymbol{e}_{i}\cdot\boldsymbol{\delta}_{ij})(\boldsymbol{e}_{s}\cdot\boldsymbol{\delta}_{ij})$ where  $\boldsymbol{S}_{i}$ and $\boldsymbol{S}_{j}$ are spin operators for the two different sites, $J_{ij}$ is the nearest neighbor magnetic exchange interaction, and  vector $\boldsymbol{\delta}_{ij}$ is connecting neighboring sites $i$ and $j$ from different magnetic sublattices. For an AF on a square\cite{Fleury:1968} or slightly distorted square lattice\cite{Vernay2007a} one expects signatures of magnetic exchange Raman scattering in a cross-polarization.   In our results, two-magnon band is present in $b(\hat c \hat a)\bar{b}$ polarization, and has zero intensity at $b(\hat c^\prime \hat a^\prime)\bar{b}$, which is consistent  with checkerboard order. It can be regarded as  an analogy to $B_{1g}$ magnetic polarization where two-magnon Raman response is observed for for square AF\cite{Devereaux:2007}.




Two-magnon excitation is predicted at about
3$J$ for a square AF\cite{Vernay2007a,Cottam} and would be shifted to lower frequencies due to magnon-magnon interactions. The observation of the broad band centered at 500~\cm\ suggests the value of  spin exchange constant of about $J$=180 \cm\ for $\kappa$-ET-Cl and $\kappa$-ET-Br. This value is close to that  estimated from  a fit of magnetic susceptibility data  $J$=250 K (174~\cm, 21.5 meV)\cite{Miyagawa:1995,Powell06,Shimizu03}. A value  proposed  based on NMR data  $J$=50meV (403~\cm)\cite{Smith2003} is twice higher, however the measurements from which the value was obtained were performed at magnetic fields above spin-flop field.

The increase of the spectral weight of magnetic background starts at about 200~K (inset in Fig.~\ref{Cl}).  On cooling the maximum reduces in width from approximately 600 to 400 \cm, and slightly shifts to higher frequencies, from 420 at 200 K to 480 \cm at 15~K. However, we do not observe any discontinuous  change  on T$_N$=23~K in Raman spectra of AF $\kappa$-ET-Cl. This is in contrast with the results on two-magnon  scattering for most of magnetic materials, including highly frustrated ones, which show a narrowing and high-frequency shift of the two-magnon features at T$_N$\cite{Cottam,Valentine2015}.
This temperature dependence suggests, that the in-plane short-range  magnetic correlations which have dominant importance in the two-magnon Raman response are fully developed by T$_N$ in $\kappa$-ET-Cl.  It is in agreement with  the heat capacity measurements for $\kappa$-ET-Cl\cite{Yamashita2010}, where the absence of any feature at T$_N$ is explained by the fact that all the entropy is released on the formation of two-dimensional AF correlations at higher temperatures.



\begin{figure}
  \includegraphics[width=9cm]{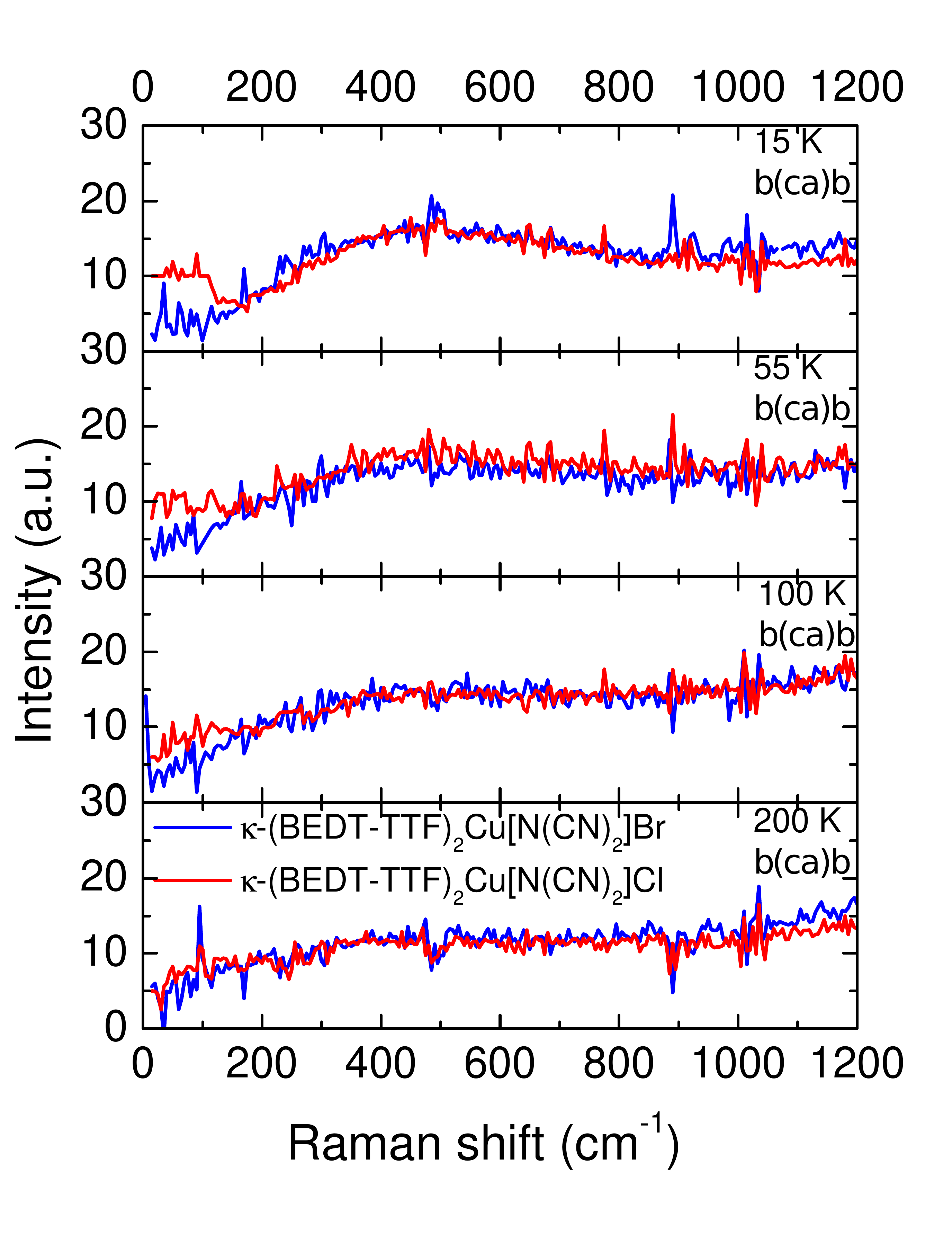}\\
  \caption{Comparison of the 15\,K Raman spectra of \etbr\ (blue lines) and \etcl\ (red lines)  at 200, 100, 55 and 15\,K. The phonon lines are subtracted. The spectra show a comparison of the two-magnon features of the two materials.
  }
  \label{Compare}
\end{figure}

The spectra of FL metal $\kappa$-ET-Br shows two-magnon feature that is very similar to AF $\kappa$-ET-Cl and appears in the same  $b(\hat c \hat a)\bar{b}$  polarization and frequency range around 500~\cm, and shows a similar temperature dependence of the spectral weight. In Fig.~\ref{Compare} we present the comparison of $\kappa$-ET-Cl and $\kappa$-ET-Br spectra in the $b(\hat c \hat a)\bar{b}$ polarization at 200, 100, 55 and 15 K after subtraction of the phonons. At each temperature, the spectra in the region of two-magnon excitations basically coincide. The difference between $\kappa$-ET-Cl and $\kappa$-ET-Br behaviour below approximately 150~\cm\  observed at 15 K is correlated with  the appearance of coherent charge carriers response at that frequency range in the conductivity spectra of $\kappa$-ET-Br  below 50 K\cite{Dumm2009}.

In the temperature range between 200  and 50~K, our results are in agreement with the results of NMR experiments, where an identical  increase of the spin-lattice relaxation rate $(T_1T)^{-1}$ is observed for these two materials due to increase of  AF fluctuations\cite{Kawamoto1995}.
Below 50~K,  $(T_1T)^{-1}$  for the Mott insulator $\kappa$-ET-Cl increases and diverges at T$_N$, while for $\kappa$-ET-Br in the  Fermi-liquid regime,\cite{Dumm2009} $(T_1T)^{-1}$  decreases with temperature. In contrast to that, in Raman scattering  we observe the similar increase of the two-magnon feature in both compounds below 50~K. Since no evidence of insulating domains in $\kappa$-ET-Br is detected, our measurements capture the intrinsic response of the FL metallic state. This demonstrates that in the whole temperature range the 2D AF correlations in the metallic state have the same strength as in the AF ordered compound. Our results suggest that the decrease of  $(T_1T)^{-1}$ in the FL regime of $\kappa$-ET-Br can be explained by the presence of a pseudogap.


It is interesting to compare our results to other  metallic materials close to Mott transition.  In cuprate superconductors  the two-magnon excitations are very prominent not only in the Raman spectra of the parent AF ordered insulators, but also in the underdoped materials \cite{Sugai:2003}, providing evidence for  AF fluctuations in these metals. Similar to the observation for $\kappa$-ET-Br, in the metallic regime in cuprates  NMR measurements show a decrease of spin-lattice relaxation rate\cite{Monien1991}.  However,  the two-magnon band widens and shifts to lower frequencies by doping from the parent AF to a non-FL metal. In contrast to that, with the precision of our measurement we observe no difference in the two-magnon band position or shape between AF ordered $\kappa$-ET-Cl and Fermi-liquid state tuned by bandwidth $\kappa$-ET-Br. The  effective number of carriers per dimer lattice site in the latter can be estimated from  optical conductivity  data to be approximetaly $N_{\rm eff}= 0.35$ [\onlinecite{Merino2008}]. A doping of this level leads to a broadening of the two-magnon band similar to that of underdoped and optimally doped cuprate superconductors.

This difference between FL state in 2D organic conductors and doped cuprates suggests that it is can be essential for the parameters of magnetic correlations in the FL state that  the transition to this  state  is achieved  by changing effective electronic correlations $U/t$, while preserving stoichiometry of the material and without introduction of disorder. The presence of a two-magnon band in the spectra of $\kappa$-ET-Br suggests the same parameters of short-range 2D AF fluctuations in the AF insulating and FL liquid state close to AF order in these compounds.


In conclusion, using Raman scattering we observe two-magnon excitations in AF ordered $\kappa$-ET-Cl as a wide band centered at about 500~\cm\ in $b(\hat c\hat a)b$ polarization, analogous to  $B_{1g}$ symmetry of square AF. The polarization dependence of this band suggests a checkerboard spin pattern in the $(ac)$ plane. The position gives an estimate of  magnetic exchange constant $J$ of about 180~\cm. We observe also a two-magnon excitation band in the Fermi-liquid metal $\kappa$-ET-Br, with the temperature, frequency and intensity dependence being the same as for AF  $\kappa$-ET-Cl. These results demonstrate that in-plane  AF correlations in the FL liquid state of  the organic superconductor $\kappa$-ET-Br have a similar strength as in AF Mott insulator. This observation is a strong case for the suggested interrelation of spin fluctuation and superconductivity in $\kappa$-ET-Br.


\section{Acknowledgements}

Authors thank R. Valenti, B. Powell, and N. P Armitage for fruitful discussions. N.D. grateful acknowledges support via the Margarete von Wrangell-Habilitationsprogramm and an M. Hildred Blewett Fellowship and thanks the Walther-Meissner-Institut  and 1. Physikalisches Institut, University of Stuttgart for hospitality. The work in JHU was supported by the US Department of Energy, Office of Basic Energy Sciences, Division of Material Sciences and Engineering under grant DE-FG02-08ER46544. R.H. gratefully acknowledges support by the DFG via the Transregional Collaborative Research Center TRR\,80 and the Bavarian Californian Technology Center BaCaTeC (project no. A5\,[2012-2]).

\bibliography{./literatureR}

\end{document}